# Engineering consensus in static networks with unknown disruptors

Agathe Bouis, Christopher Lowe, Ruaridh A. Clark, and Malcolm Macdonald

Centre for Signal and Image Processing (CeSIP), Department of Electric and Electrical Engineering, University of Strathclyde, Glasgow, Scotland

agathe.bouis@strath.ac.uk

## Abstract

Distributed control increases system scalability, flexibility, and redundancy. Foundational to such decentralisation is consensus formation, by which decision-making and coordination are achieved. However, decentralised multi-agent systems are inherently vulnerable to disruption. To develop a resilient consensus approach, inspiration is taken from the study of social systems and their dynamics; specifically, the Deffuant Model. A dynamic algorithm is presented enabling efficient consensus to be reached with an unknown number of disruptors present within a multi-agent system. By inverting typical social tolerance, agents filter out extremist non-standard opinions that would drive them away from consensus. This approach allows distributed systems to deal with unknown disruptions, without knowledge of the network topology or the numbers and behaviours of the disruptors. A disruptor-agnostic algorithm is particularly suitable to real-world applications where this information is typically unknown. Faster and tighter convergence can be achieved across a range of scenarios with the social dynamics inspired algorithm, compared with standard Mean-Subsequence-Reduced-type methods.

## Introduction

Methods of network control are shifting from centralised to distributed to avoid communication overheads, their associated latencies and single point failures [1, 2]. Distributed control strategies must cope with issues of security and resilience to faults, and deliberate attacks against ever growing networks (both in number of nodes and connections). Although such issues are also present in centralised systems, distributed networks prove to be particularly vulnerable as they lack a central authority to authenticate nodes as well as monitor and detect misbehaviours or failures [3]. Decentralised control methods are thus characterised by the realistic systems to which they are applied: systems with unknown numbers of disruptors and unknown network topologies. Typical networks include the likes of Internet of Things (IoTs), mobile robotic systems, smart power grids, and wireless sensor networks.

The foundation of distributed control is consensus, whereby nodes within a system reach a common shared state [4]. This is often achieved using linear consensus iterations, where at each time step network nodes update their state values based on a weighed combination of their neighbours' values [5, 6]. The choice of combination is what decides on a consensus' disruption tolerance. Disruption tolerance is commonly defined as the ability to cope with, and/or recover from disruptions within the network [7]. These malfunctions can be faults or deliberate attacks [8], which are manifested in this paper under the same form: undetectable disruptions. The simplest approach to selecting which neighbouring nodes to combine is mean-based computation; where the average of all neighbours' values is used for the state update. However, these types of methods are not resilient as nodes can be swayed by the presence of disruptors [8]. They must therefore be adapted to ensure fault-resilience; usually by filtering incoming neighbour state values and/or by augmenting the system in some way such as introducing virtual connectivities to collect more values on which to base the update [9], or implementing observers to create adaptive controllers [10, 11]. One such typical approach is the Mean-Subsequence-Reduced (MSR)-type method [12], and its extensions (e.g. Weighted-MSR [8, 13], Omission-MSR [14], …). MSR methods ensure their fault-tolerance by removing a set number of the largest and smallest values received by nodes at each consensus step, enabling agents to reach an agreement despite disruptive values. The removed numbers correspond to the maximum possible number of faults that the system can cope with [15]. By requiring knowledge of this design value, MSR method are observed to respond poorly to situations where this information is unavailable as well as lacking robustness to additional failures.

To provide a dynamic and robust solution to the consensus problem, requiring neither information regarding the topology nor indication of the number of disruptive agents in the system, this paper presents a fully decentralised, adaptive consensus algorithm handling the specifications of realistic, static networks using a framework adapted from the dynamics of social systems. Specifically, a sociophysics approach is adopted. Sociophysics models the behaviour of human crowds and the social interactions therein [16, 17]. That is, people in everyday life influence one another and adapt towards, or away from, the expressed opinions of the people they have interacted with.

A model of interest, notably regarding the spreading of minority opinion [18], is that proposed by Deffuant et al [19]. Deffuant's Bounded confidence (BC) is a stochastic model of the evolution of continuous-valued opinions within a given range [20], usually set between -1 and 1 [21]; moderate opinions close to 0 and extreme ones approaching either -1 or 1. The framework can be understood as: two agents coming into contact must hold sufficiently close opinions in order to mutually influence one another. When encountering agents of radically different opinions, that is, when the difference of opinions is too large, there are no changes with respect to the opinions held prior to the communication. The threshold for the difference is known as the agent's tolerance, $d$. Denoting the opinion of agent $i$ at a given time $t$ as $o_i[t]$, the interactions of agent $i$ with agent $j$ can be described as

$$o_i[t + 1] = o_i[t] + \eta(o_j[t] - o_i[t]) \text{ if } |o_j[t] - o_i[t]| < d .$$

Eq. 1

When implementing a fixed universal tolerance, the final distribution of opinions can be categorised as; consensus, when a single opinion dominates; polarisation, when two opinions emerge; and, fragmentation, when several opinion clusters can be observed [9]. The transition from one phase to the next depends on the universal tolerance threshold.

A secondary dimension to improve the model's realism was introduced by Sobkowicz [22]. By deriving an explicit correlation between extreme opinions and low tolerance, the tendency for extremists to be inflexible is explained. It also illustrates the propensity of moderate agents to be more open minded and thus easily swayed towards extremism [22-24]. This can be exploited by bad actors to lead to dissention, polarisation, and fragmentation of opinions [23, 25, 26]. This behaviour is addressed by morphing the universal tolerance, $d$, into individual agent's tolerance. These tolerances are then updated as a function of the agent's own opinion with respect to other opinions as $d_i = f(|o_i|)$.

The mechanism of belief diffusion is not solely observed in opinion dynamics, it also represents the foundations of consensus algorithms in network wherein agents communicate with their (usually one-hop) neighbours to update their own state variables. The source of conflict in both model and algorithm can be understood to stem from similarly disruptive attitudes. As observed by the rise in political extremism propelled by small but vocal minorities, in social environments agents with extreme opinions compel other non-extreme agents (moderates) into becoming extremists themselves [16, 18]. This is also the case for engineered networks, where the presence of even small minorities of disruptive, selfish, or malicious agents can come to prevent consensus [3].

This paper exploits, and inverts, the mechanisms of opinion diffusion to design a consensus algorithm resistant to faulty/disruptive actors (extremists). The algorithm, termed the Opinion Dynamics-inspired DIsruption-tolerant Consensus, ODDI-C (pronounced *o·duh·see*), enables a resilient consensus to be reached despite a lack of awareness of network topology and number of disruptive agents. This approach is chosen based on the similarity between how agents in bounded confidence social models and nodes in linear consensus scenarios update their opinions/values. On the basis of these similar mechanisms of opinion/state value update, features of social systems resulting in consensus are analysed, extracted, and implemented in the context of a consensus algorithm. In practice, this is achieved through the exploitation of a dynamic tolerance linked to extremism, whereby agents filter out extreme non-standard opinions driving them away from consensus. This make the approach particularly suitable to real-world applications where there is limited, to no, external information and where synchronisation is critical.

## Opinion Dynamics-inspired DIsruption-tolerant Consensus (ODDI-C) Algorithm

The ODDI-C Algorithm adapts the concept of tolerance from social systems, using an analogous approach to the Deffuant Model for its update and filtering. That is, an agent's tolerance increases when it finds itself more extreme, and all values beyond an agent's tolerance are filtered out and not considered during the update. Moreover, the ODDI-C Algorithm deals with the median-based z-scores rather than raw opinions $\psi_i[t]$. The median-based z-score quantifies how close an opinion is to the local median of the distribution. The replacement of the raw opinion with the z-score serves two purposes. The first is to allow for any range of input to be considered, thus expanding the typical Deffuant model's range of -1 to 1. Secondly, and more importantly, the use of the z-score serves as inbuilt tolerance indicator.

The consensus of compliant agents can be ensured by having them only consider opinions which are closer to the median than they are. The z-score facilitates this by acting as a dynamic tolerance for the agent doing the filtering, enabling it to only consider opinions whose z-score are smaller than its own. Put otherwise, agents standing close to the median of the distribution will prove particularly stubborn while agents who are outlier themselves will be much more willing to account for other agents' opinions. This is an inversion of typical social systems, for whom extremism is linked to low tolerances. This approach also provides an elegant solution to the problem of filtering out extreme values while allowing for a level of adaptability dependent on the agent's own opinion.

Note here the use of the median rather than the mean when dealing with z-scores due to its statistical robustness to outliers. The median is more stable than the mean when dealing with heavy tailed or asymmetric distributions, and as such, it is a more appropriate tool for outlier filtering [27]. Adapted from [28], three consensus properties act as algorithm design requirements.

- Termination: All compliant agents must decide on a value.

- Validity: The agreed upon value is not influenced or dictated by disruptive agents.

- Agreement: No two compliant agents decide on different values, to within an agreed tolerance.

Note that Termination does not require Agreement, in which case polarisation or fragmentation are obtained.

The ODDI-C Algorithm is primed for systems dealing with faults/attacks of unknown identities and behaviours Considering this unknown, the validity property is impossible to ensure [29]. The property is thus weakened, following [29], to be applicable for such scenarios, which use linear consensus mechanics with unknown disruptors.

- Weak Validity: If all nodes are compliant and propose the same value, all nodes will agree on this value.

**Algorithm Rules**

The algorithm assumes a network constructed as follows: Let $G(t) = (V, \varepsilon, A)$ be a digraph with a finite number of nodes (agents) $V = 1, 2, \ldots, N$, a set of directed edges $\varepsilon \subseteq V \times V$, and an adjacency matrix $A = a_{ij} \in R(N \times N)$. Each directed edge $(j, i) \in \varepsilon$ represents a directional link between the node pair $(j, i)$, such that communication between the nodes is enabled. In the model described here, it is assumed that the nodes communicate synchronously at each given opportunity such that the edges model the flow of information between the two nodes. The set of nodes influencing node $i$, it's in-neighbours, is defined $V_i^{in} = j \in V : (j, i) \in \varepsilon$ while the nodes influenced by $i$, its out-neighbours, are in the set $V_i^{out} = j \in V : (i, j) \in \varepsilon$. The components of the adjacency matrix $a_{ij} = 1$ if $j \in V_i$, that is, if node $i$ can receive information from node $j$, otherwise $a_{ij} = 0$. It should also be noted that each node has access to its own value at each time step, thus making it consider its inclusive neighbourhood.

The synchronous ODDI-C Algorithm implemented by compliant agents is presented in Figure 1.

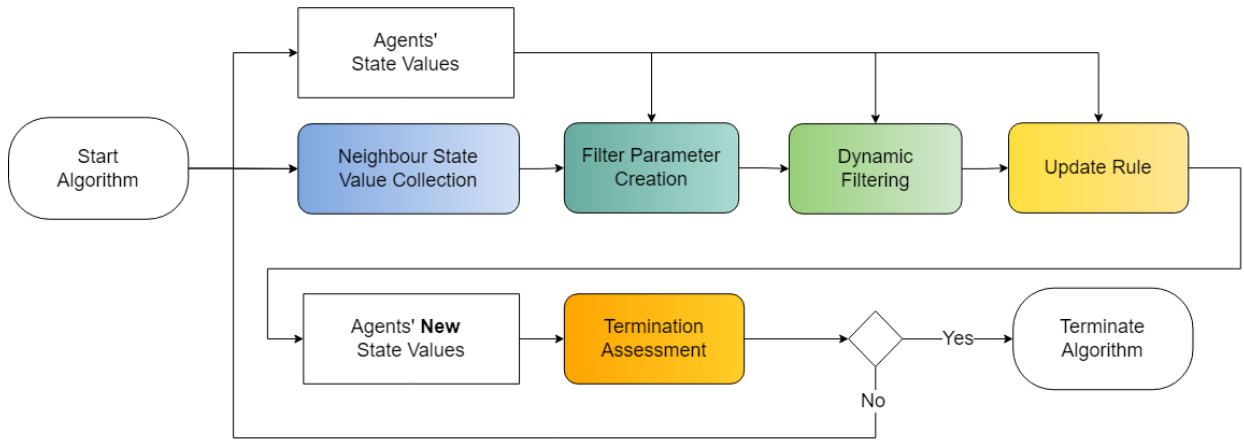

*Figure 1: ODDI-C Algorithm Flow Diagram. The algorithm is repeated for each compliant node at each time step until the termination criterion (or criteria) is (are) met.*

The algorithm's filtering process works as follows, starting with **Neighbour State Value Collection** (see Figure 1). At each time step $t \in N$, each compliant node $i$ sends its state variable $\psi_i[t]$ to all of the nodes in its out-neighbourhood $j \in V^{out}$ out and in turn receives the state variables $\psi_j[t]$ from its in-neighbours $j \in V^{in}$ (compliant and disruptive, if they exist). These values are known as $\Psi_i[t]$,

$$\Psi_i[\mathrm{t}] = \psi_j[\mathrm{t}] : j \ \in V_i^{in} \ . \qquad \text{Eq. 2}$$

From these received values, the **Filter Parameter Creation** can begin. These parameters are the median $\widetilde{\Psi}_i^{\psi_i}[t]$ and the Median Absolute Deviation $\mathrm{MAD}_i^{\psi_i}[t]$ (a more robust alternative to the standard deviation) of the concatenated array $\Psi_i[t]$ and are determined from

$$\Psi_i^{\psi_i}[t] = \mathrm{median}(\Psi_i[t]) \qquad \text{Eq. 3}$$

$$\mathrm{MAD}_i^{\psi_i}[t] = \mathrm{median}\big(\big|\Psi_i[t] - \widetilde{\Psi}_i^{\psi_i}[t]\big|\big) \ . \qquad \text{Eq. 4}$$

When dealing with normal distributions, the MAD is scaled using a factor of 1.4826 [27, 30]. The use of the MAD and its normalised scaling,

$$\mathrm{NMAD}_i^{\psi_i}[t] = 1.4826 \cdot \mathrm{MAD}_i^{\psi_i}[t], \qquad \text{Eq. 5}$$

relies on the assumption of a normal mostly symmetric distribution of values. For other distribution, other estimators or scalings may be more appropriate [30]. This in turn allows for the median-based z-scores of the received values to be calculated. These values are kept as absolutes as the sign is of no importance for the filtering. The z-scores of the received values is determined as,

$$\text{z-score}_i(\Psi_i[t])[t] = \left( \frac{\big|\Psi_i[t] - \widetilde{\Psi}_i^{\psi_i}[t]\big|}{\mathrm{NMAD}_i^{\psi_i}[t]} \right), \qquad \text{Eq. 6}$$

using the median ($\Psi_i^{\psi_i}[t]$) and scaled MAD ($\mathrm{MAD}_i^{\psi_i}[t]$) of the distribution. The filter value,

$$\mathrm{filter}_i^i[t] = \left( \frac{\big|\psi_i[\mathrm{t}] - \widetilde{\Psi}_i^{\psi_i}[t]\big|}{\mathrm{NMAD}_i^{\psi_i}[t]} \right)$$
$$= \text{z-score}_i^i, \qquad \text{Eq. 7}$$

is then created based on the z-score of the agent (and its opinion $\psi_i[t]$).

Before the filter can be used, it must be optimised. This is accomplished through the implementation of dynamic upper limit. The method used is a Hampel filter, a method typically applied to limit the classification of nodes as outlier/not-outlier. Nodes are classified as outliers if they are larger/smaller than 3 scaled Median Absolute Deviations (NMAD) [27]. Considering our use of z-score as discriminant (rather than the raw state value), the Hampel filter is adapted as:

$$\mathrm{filter}_{max}[t] = 3 \qquad \text{Eq. 8}$$

$$\text{if} \quad \mathrm{filter}_i^i[t] \geq \mathrm{filter}_{max}[t] \quad \text{then} \quad \mathrm{filter}_i^i[t] = \mathrm{filter}_{max}[t]. \qquad \text{Eq. 9}$$

Once the filter value is set for each agent, they can then proceed to the **Dynamic Filtering**, assessing which values will be used to contribute to the update. Values are filtered out if their z-scores are higher than the dynamic filter value. Selected values ($F_i[t]$) can be understood as

$$F_i[t] = \Psi_i\big(\text{z-score}_i(\Psi_i[t])[t] < \mathrm{filter}_i^i[t]\big) \qquad \text{Eq. 10}$$

meaning that the only values the node will accept will be closer to the median (centre of the distribution) than its own value.

Following the filtering, agents will follow the **Update Rule** to update their state values ($\psi_i[t+1]$) based on the average difference between the selected values ($f_i \in F_i[t]$) and its own past value ($\psi_i[t]$). The extent of the update will be tuned by the learning rate of the consensus $\eta$, here set to be 0.5 (refer to damped Newton's method for details). The resulting new state value is expressed as

$$\psi_i[t+1] = \psi[t] + \eta \cdot \frac{1}{|F_i|} \sum_{j=1}^{|F_i|} (\psi_i[t] - f_i[t]).$$  Eq. 11

The algorithm is repeated for each compliant node and for each time step until the **Termination Assessment** is met. This assessment's criterion can be a maximum time, where the simulation stops once maximum time is exceeded ($t > t_{max}$), or a value-based decision, where the simulation stops once convergence is within acceptable threshold.

**Convergence Metric**

To visualize the evolution of compliant nodes' opinions over time, a simple metric assessing convergence is implemented. Convergence is the process by which the agreement and termination properties are met. The Convergence Metric ($CM$) is calculated at each time step as the Total Difference ($TD$) between compliant agents, normalised with respect to the initial difference between all nodes ($TD[0]$),

$$CM[t] = \frac{TD[t]}{TD[0]}$$  Eq. 12

$$TD[t] \sum_{k=1}^{V} \sum_{m=1}^{V} (|\psi_k[t] - \psi_m[t]|),$$  Eq. 13

where $\psi_k$ are agents' opinions $\psi_k \in \Psi_k$ for k $\in$ 1, V and $\psi_m$ are agents' opinions $\psi_m \in \Psi_m$ such that between the two loops k and m, all agent opinions are compared, and their absolute difference tabulated. The relative metric decreases when nodes converge, agreeing on a shared state, and increases when nodes diverge.

As the convergence metric is a relative value dependent on the initial difference between nodes' values ($TD[0]$), it can be problematic to understand on its own, even when contextualised by the plot of nodes' opinion trajectories. To avoid confusion, an error threshold ($err$) is applied. This threshold is used to create a minimum floor value ($Floor$) for the convergence metric. Results smaller than this floor are levelled to the floor value. This adaptation of the convergence metric prevents the development of small rounding errors while giving context to the results.

$$Floor = \frac{err}{TD[0]}$$  Eq. 14

$$\text{if} \qquad CM < Floor \qquad \text{then} \qquad CM = Floor,$$  Eq. 15

the error threshold ($err$) is set at a value of $10^{-7}$.

# Analysis

The convergence of the consensus algorithm is proven by assessing the changing (converging) boundedness of compliant agents at each time step. The update of the algorithm, from Eq. 11, which can be written as

$$\psi_i[t+1] = \psi_i[t] + \epsilon \cdot \mathbf{1}_{F_i}^T \cdot \left( \psi_i[t] \cdot \mathbf{1}_{F_i}^T - \frac{1}{|F_i|} F_i[t] \right).$$

<div align="right">Eq. 16</div>

This can be rearranged as

$$\psi_i[t+1] = \psi_i[t] + \epsilon \cdot \mathbf{1}_{F_i}^T \cdot \left( \psi_i[t] \cdot \mathbf{1}_{F_i}^T - \mu_i^{F_i}[t] \right)$$

<div align="right">Eq. 17</div>

where $\mu_i^{F_i}[t]$ is the local mean of the filtered values $F_i[t]$ at time $t$. It should be noted that $\mu_i^{F_i}[t]$,

$$\mu_i^{F_i}[t] = \frac{1}{|F_i|} F_i[t],$$

<div align="right">Eq. 18</div>

differs from $\mu_i[t]$ (the mean of all values received by agent $i$) as extremist values are filtered out of the distribution. Given this rearranging,

$$\lim_{t \to \infty} \psi_i[t] \to \mu_i^{F_i}[t] .$$

<div align="right">Eq. 19</div>

As established by the algorithm rules, the values $F_i[t]$ change at each time step and so will the local filtered mean ($\mu_i^{F_i}[t]$). The local filtered mean will in turn evolve to a global mean $M[t]$ as the agents communicate with each other and disruptions are filtered out. Although not a static value, the global mean will remain stable for the most part, being a weighted mean of the local means, in addition to being a function of the network topology and of the initial state values

$$\lim_{t \to \infty} \mu_i^{F_i}[t] \to M[t]$$

<div align="right">Eq. 20</div>

$$\therefore \lim_{t \to \infty} \psi_i[t] \to M[t].$$

<div align="right">Eq. 21</div>

This can be rearranged such that

$$\therefore \lim_{t \to \infty} (\psi_i[t] - M[t]) \to 0,$$

<div align="right">Eq. 22</div>

which can be understood in terms of the bounds as

$$|\psi_i[t+\Delta] - M[t+\Delta]| \le |\psi_i[t] - M[t]|$$

<div align="right">Eq. 23</div>

where $t + \Delta > t$ as $\Delta > 0$ is a positive non-zero value representing a change in time. Note that the consensus algorithm operates in a discrete synchronous manner. On the basis that all agents send (and receive) their opinions at the same time and at the same frequency, time can be counted in a discrete fashion with time steps. As such, $\Delta$ can also be viewed as a non-zero number of time steps. When looking at the state values $\psi[t]$ of all the compliant agents

$$\max(\psi_i[t+\Delta]) \le \max(\psi_i[t])$$

<div align="right">Eq. 24</div>

$$\min(\psi_i[t+\Delta]) \ge \min(\psi_i[t]).$$

<div align="right">Eq. 25</div>

Such that if $r$ is the range of opinions of compliant nodes

$$r[t] = \max(\psi_i[t]) - \min(\psi_i[t])$$

<div align="right">Eq. 26</div>

$$r[t+\Delta] = \max(\psi_i[t+\Delta]) - \min(\psi_i[t+\Delta])$$

<div align="right">Eq. 27</div>

$$r[t + +\Delta] \leq r[t] \ .$$ Eq. 28

This means that a decreasing range of agents' opinions is observed over time as the agents converge. While it is known that the agents will tend towards their local mean, as a result of the filtering mean-based consensus, the "pull" or attraction towards the local mean may lead an agent's opinion further away from the global mean. However, if sufficient exchange of data takes place, such as the ones ensured by meeting the sufficient and necessary conditions of $[(F + 1, F + 1) - robustness]$ (where *robustness* is described by LeBlanc in [4]), this local mean converges to the global mean.

**Numerical Experimentation**

Four numerical experiments are used to investigate the algorithm's performance. Experiment 1 considers a single network with disruptors, where the performance of the ODDI-C Algorithm is compared against a scenario with a classical mean-based algorithm (no filtering) implemented. This allows for the confirmation of the Weak Validity property and for the behaviour of compliant nodes following ODDI-C to be examined in the face of specific patterns of disruptive behaviours. Experiment 2 assesses the performance impact of changing network connectivity in the face of a constant number of disruptions using a series of small Monte Carlo simulation (50 runs) per connectivity level. This test enables for ODDI-C's performance to be measured for various connectivity levels, assessing the connectivity's impact on the Termination and Agreement properties. ODDI-C's output is compared against the MSR-Algorithm's [8] for the same networks, its behaviour being a point of reference in the field of disruption tolerant consensus (details of its implementation are provided in Section: MSR Implementation).

Experiment 3 uses a sequence of small Monte Carlo simulations to evaluate ODDI-C's performance degradation (deterioration with the Termination and Agreement properties) when dealing with increasing numbers of disruptions for networks with fixed connectivity. This test is used to investigate the Termination and Agreement properties of the algorithm, particularly regarding $(|D|)$, the number of disruptors, which overwhelm attempts at consensus building. Similar, to Experiment 2, each simulation is duplicated using MSR as comparison. Experiments 2 and 3 are both performed on 20-node networks to allow for a sufficient number of message exchanges and nodes to showcase the algorithm's scaling while remaining of a comprehensible and manageable size.

The premise of ODDI-C Algorithm is an undetectable fault or attack scenario disrupting the system's ability to achieve Termination and Agreement [31]; meaning, disruptions inhibit the Weak Validity property. Disruptive agents are understood as being "deceptive", with non-disruptive agents remaining unaware of their presence in the system. Disruptive agents do not follow the ODDI-C Algorithm, instead adopting unknown behaviours and/or values and broadcasting them [32]. Three types of disruptions are used across the three Experiments: a sine wave (T1, described in Eq. 29), a linearly increasing function (T2, described in Eq. 30), and a noise pattern (T3, described in Eq. 31). The functions describing them are

$$\left(\overline{\Psi}[0] + s_y\right) + (A \cdot r) * \sin(\omega t + s_x)$$ Eq. 29

$$\overline{\Psi}[0] + m \cdot t$$ Eq. 30

$$\overline{\Psi}[0] + \text{rand} \cdot 0.9 \cdot r,$$ Eq. 31

where, $\overline{\Psi}[0]$ is the average of all nodes (including ones that will become disruptive) at time 0, $s_y$ is the vertical y-axis shift of the sine wave disruption, $s_x$ is the horizontal x-axis shift of the sine wave disruption, $A$ is the amplitude factor of the sine wave disruption, $r$ is the range of node values at time 0, $\omega$ is the angular frequency of the sine wave disruption, $m$ is the gradient of the linear function, and rand is a uniformly distributed random number in the interval [0,1]. Initial node values of are drawn from normal distributions with mean $\mu$ and standard deviation $\sigma$, and described as

$$\Psi[0] \sim \mathcal{N}(\mu, \sigma^2).$$

<div align="right">Eq. 32</div>

In all test cases, disruptive agents become active from time step 2 onward. These nodes keep the same identity throughout the run. The nodes do not recover and are considered to remain disruptive for the extent of each run. The convergence metric is determined and used to compare ODDI-C and MSR's behaviour in multi-run scenarios. The termination criterion for all Experiments is a maximum time $t_{max} = 40$, with the simulation stopping once this maximum time is reached.

For Experiment 1, the network robustness is provided. This metric is provided to enable comparisons of the proposed algorithm performance to other consensus algorithms, which test their algorithms on graphs described by their robustness. Networks are changed randomly for Experiments 2 and 3, as such their robustness are not provided considering their lack of relevance. As defined by LeBlanc [8], network robustness formalises the notion of redundancy of direct information exchange between the various subsets in the network. It is a term which captures the idea that there are enough nodes in every pair of non-disjoint sets within the network that have at least a given number of neighbours outside of their respective sets. That is, that each node receives enough information so that it will not be swayed by disruptive agents attempting to "contaminate" the subset to which the node belongs. The reader is directed towards [8] for details regarding this robustness calculation.

## MSR Implementation

MSR relies on knowledge of the number of disruptions $|D|$ in the system, with at most $|D| \times 2$ being filtered out. At each time step, each node $i$ receives its in-neighbours' values ($\Psi_i[t]$), forms a sorted list, and proceeds to the filtering. If there are less than $|D|$ values larger than the node's own value ($|D| > |\Psi_i[t] > \psi_i[t]|$), then node $i$ only removes the values strictly larger than its own ($|\Psi_i[t] > \psi_i[t]|$), otherwise it removes precisely the number of disruptive values known to be present in the system ($|D|$). Likewise, if there are less than $|D|$ values strictly smaller than the node's own value ($|D| > |\Psi_i[t] < \psi_i[t]|$), only this number will be removed. Otherwise, precisely $|D|$ smaller than the node's values will be removed. The remaining values $F_i[t]$ are used in linear consensus update as described in Eq. 11. MSR's implementation in Experiments 2 and 3 follows [8].

## Experiment 1: Single run

Two simple scenarios are first considered, focusing on assessing single runs on two networks. Two multi agent networks are assessed, one with $N = 7$ nodes and the other with $N = 15$ nodes. In the first case, the disruptive agents are $D = 1, 7$ and in the 15-node case, they are $D = 4, 12$. The 7- node network is (3,3)-robust while the 15-node is (5,1)-robust (in addition to being (4,2)- and (3,2)-robust). All agents (compliant and disruptive) have initial values drawn from a normal distribution. The 7-node case have initial conditions between 10 and 100 (distribution described in Eq. 32, parameters $\mu = 50.5$, $\sigma = 24.75$ ) and deals with two disruptive sine waves (T1, as described in Eq. 29), one with a large period and amplitude, and another with a small period and amplitude. The 15-node case has initial conditions set between 1 and 15 (distribution described in Eq. 32, parameters $\mu = 8$, $\sigma = 7$) and its disruptive behaviours are a linearly increasing function (T2, as described in Eq. 30) and a noise function (T3, as described in Eq. 31). The parameters describing the disruptive behaviours in Experiment 1 are shown in Table 1 for the 7-node and 15-node cases:

*Table 1: Experiment 1 Disruptive Node Parameters*

| | $A$ | $s_x$ | $s_y$ | $\omega$ | $m$ |
|---|---|---|---|---|---|
| $D1_{\text{EXP1:7}}(T1)$ | 0.4523 | 1.5011 | 0.4289 | 1.8711 | — |
| $D2_{\text{EXP1:7}}(T1)$ | 0.4512 | 3.2642 | -1.3117 | 0.2761 | — |
| $D1_{\text{EXP1:15}}(T2)$ | — | — | — | — | -0.1707 |
| $D2_{\text{EXP1:15}}(T3)$ | — | — | — | — | — |

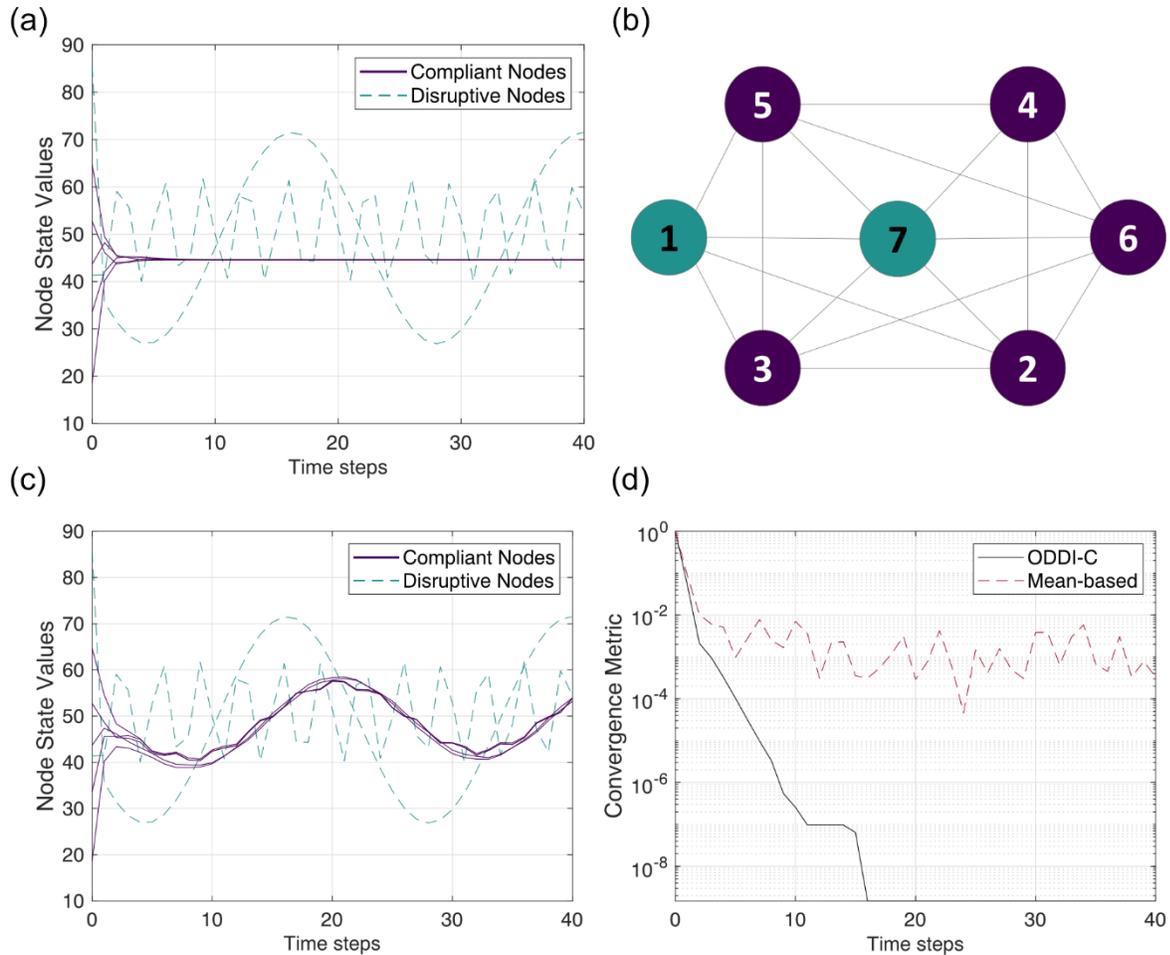

*Figure 2: Comparison of the ODDI-C and the Simple Mean-based Algorithm performances for the 7- Node (3,3)-robust Network with 2 Disruptive Agents. The two disruptive agents both exhibit a sine wave (T1, as described in Eq. 29), each with a different amplitude and period. Tiles (a and c) show the nodes' Opinion Trajectories for the ODDI-C (a) and the Mean-based algorithms (c), tile (b) shows the network diagram (compliant agents shown in purple) and tile (d) compares the scenarios' Convergence Metric (calculated as the normalised difference between compliant nodes).*

The results of the simulation of the 7-node network are shown in Figure 2; note the convergence metric's floor value of $\sim 10^{-8}$. The disruptive behaviours and the nodes' initial conditions are the same for both the ODDI-C and the Mean-based Algorithm cases. As seen in Figure 2 (a), with ODDI-C implemented, consensus is rapidly and effectively reached - unlike the case with the mean-based algorithm (c). This is expressed more clearly looking at the convergence metric, Figure 2 (d), which decreases for ODDI-C's run as compliant nodes come together and the difference between their state value reduces. This is not true for the mean-based algorithm simulation for which the convergence metric fluctuates along with the disruptive nodes' values.

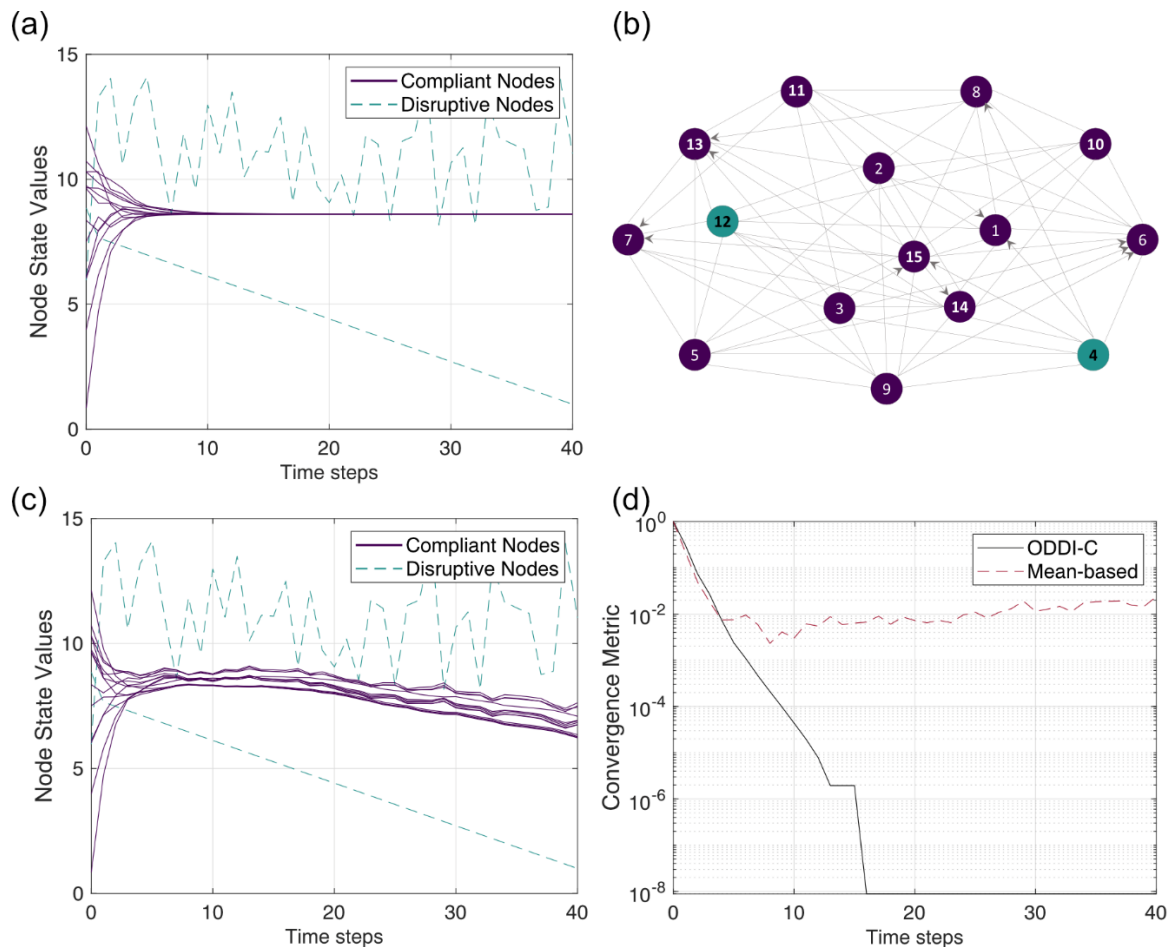

*Figure 3: Comparison of the ODDI-C and the Simple Mean-based algorithm performances for the 15-Node (3,2)-robust Network with 2 Disruptive Agents. The two disruptive agents exhibit are a noise pattern (T3, as described in Eq. 31) and a linear behaviour (T2, as described in Eq. 30) respectively. Tiles (a and c) show the nodes' Opinion Trajectories for the ODDI-C (a) and the Mean-based algorithms (c) respectively, tile (b) shows the network diagram (compliant agents shown in purple) and tile (d) compares the scenarios' Convergence Metric (calculated as the normalised difference between compliant nodes.*

The results of the simulation of the 15-node network are shown in Figure 3. Here again, the difference between ODDI-C and the Mean-based Algorithm runs are clear - with ODDI-C maximum convergence (floor value of the convergence metric) is reached in around 15 steps, while with the mean-based algorithm, the state values of the compliant nodes wobble and proper consensus is not reached.

## Experiment 2: Fixed number of disruptors, Increasing network connectivity

A twenty-node network dealing with a fixed number of disruptive nodes, $|D| = 5$, is considered. The algorithm's performance is assessed when increasing the network connectivity (in-degree) in steps of two, $V^{in} = [3, 5, \ldots, 15]$. To account for the random nature of all consensus algorithms and of possible behaviours,

Monte Carlo simulations of 50 runs per in-degree are performed. The results of the runs, meaning their convergence metrics, are averaged to showcase the trends of the simulations with error bars provided to show maximum and minimum values of each batch's convergence metrics. These are also used for the ODDI-C Algorithm's comparison with the MSR.

To allow for comparison, the behaviour of the disruptions is kept constant across runs and Monte Carlo batches ($50 \times 8 = 400$ simulations). The opinion trajectories are calculated for the first simulation, stored, and then re-used for subsequent simulations. Note however that while the behaviours are fixed, the identities of the nodes exhibiting this disruptive behaviour are not, being selected at random from the pool of all node for each run. The disruptors are three sine waves (T1, described in Eq. 29), a linear function (T2, described in Eq. 30), and a noise function (T3, described in Eq. 31). The parameters describing the disruptive behaviours are shown in Table 2.

*Table 2: Experiment 2 Disruptive Node Parameters*

|  | $A$ | $s_x$ | $s_y$ | $\omega$ | $m$ |
|---|---|---|---|---|---|
| $D1_{\text{EXP2}}$(T1) | 0.1137 | 1.2050 | -3.9058 | 3.1381 | — |
| $D2_{\text{EXP2}}$(T1) | 0.5181 | 0.9961 | -1.5561 | 0.2217 | — |
| $D3_{\text{EXP2}}$(T2) | — | — | — | — | 0.5738 |
| $D4_{\text{EXP2}}$(T1) | 0.39528 | 4.5602 | -0.0241 | 0.2203 | — |
| $D5_{\text{EXP2}}$(T3) | — | — | — | — | — |

For each run, a new uniformly random graph is constructed. Taking as input the batch's in-degree, each node's in-neighbours are selected uniformly at random, without replacement, from the vector of all other nodes in the system. This simple random selection of in-degrees allows for a maximum variation between the tested digraph networks and their resulting adjacency matrix.

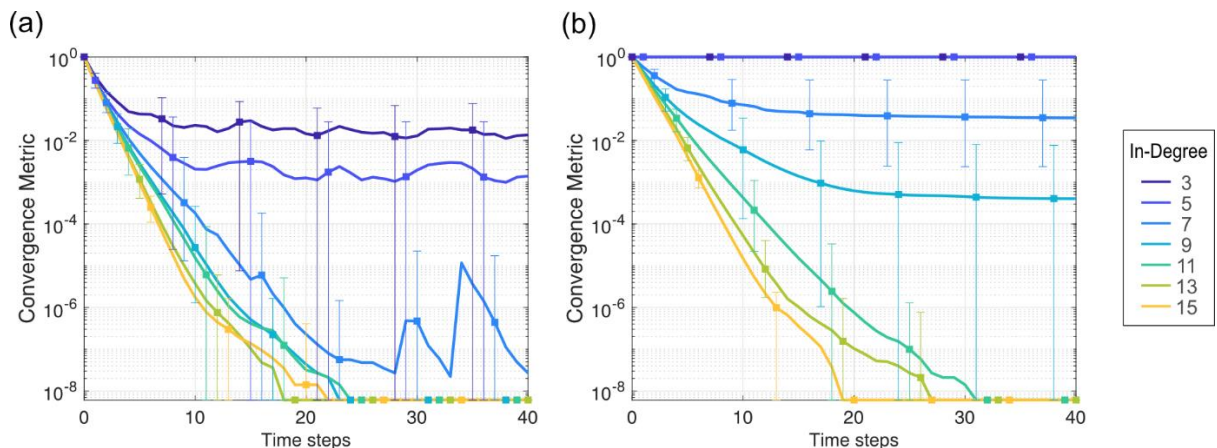

*Figure 4: Comparison of the ODDI-C (a) and MSR (b) Algorithms' convergence metrics.*

The ODDI-C Algorithm's outperforms the MSR (a pattern repeated for other disruptive behaviours) as observed across Figure 4. For each connectivity, ODDI-C reaches consensus with equal or better performance than MSR.

When simulating high in-degrees, ODDI-C reaches consensus with equal or fewer time steps than MSR despite having no knowledge of the number of disruptive agents within the system. Lower in-degree exhibit similar behaviour, due to the high ratio of disruptor nodes to connectivity, the MSR filters out most if not all values received by nodes. As a result, compliant nodes cannot reach any form of consensus (see tile (b) of Figure 4). This is not the case of ODDI-C, which even with low in-degrees manages to bring nodes to some form of low-level consensus in the network.

## Experiment 3: Fixed network connectivity, Increasing number of disruptors

The scenario investigated looks at a twenty node network with a fixed connectivity (in-degree of 6 for all agents) dealing with an increasing number of unknown disruptors, $|D|$, starting with no disruptors and increasing in steps of 1 until reaching a maximum of 8: ($|D| = [0, 1, \dots, 8]$). Considering the random nature of all consensus algorithms and of their possible behaviours, short Monte Carlo simulations of 50 runs are performed for each batch of disruptors. The results of the runs, meaning their convergence metrics, are averaged to showcase the trends of the simulations, with error bars showing maximum and minimum convergence metric values being shown at constant intervals. These are also used in the comparison with the MSR.

To enable comparison between the ODDI-C and MSR algorithms, the behaviour of the disruptions is kept constant across runs, and Monte Carlo batches. For the first run of the second simulation batch (number of Disruptive agents ($|D|$ = 1)), the opinion trajectory of the disruptive node is calculated before being stored. This opinion at each time step is then re-used for the other 49 runs of the batch ($|D|$ = 1) as well as being used by later batches with more disruptive agents. Taking the case of $|D| = 7$, the behaviour of D=1 is the same as for $|D| = 2, 3, \dots, 8$. This continuity enables for a fairer comparison of not only runs, but also MSR against ODDI-C Algorithm. Note that the node identity of the disruptors is non-constant, being selected at random from the pool of all node for each run. This Experiment's disruptions are four sine waves (T1, described in Eq. 29), a linear function (T2, described in Eq. 30), and three noise functions (T3, described in Eq. 31). The parameters of these disruptions are shown in Table 3.

*Table 3: Experiment 3 Disruptive Node Parameters*

|  | $A$ | $s_x$ | $s_y$ | $\omega$ | $m$ |
|---|---|---|---|---|---|
| D1$_{\text{EXP3}}$(T1) | 0.3666 | 1.6825 | 0.2100 | 1.2452 | — |
| D2$_{\text{EXP3}}$(T1) | 0.08841 | 4.3958 | 1.1349 | 1.12364 | — |
| D3$_{\text{EXP3}}$(T3) | — | — | — | — | — |
| D4$_{\text{EXP3}}$(T1) | 0.4889 | 0.4714 | -0.0241 | 4.4269 | — |
| D5$_{\text{EXP3}}$(T3) | — | — | — | — | — |
| D6$_{\text{EXP3}}$(T3) | — | — | — | — | — |
| D7$_{\text{EXP3}}$(T1) | 0.6072 | 3.0984 | -2.0524 | 2.4376 | — |
| D8$_{\text{EXP3}}$(T2) | $\mid$ | $-$ | $-$ | $-$ | $-$ | -0.3156 |

The results of Experiment 3 are shown in Figure 5. The convergence metrics are plotted for all investigated disruptor numbers, with error bars. As before, ODDI-C sees a performance equal or better than that of the MSR in all cases. For cases with no or low numbers (0-3) of disruptive agents, the convergence patterns are the similar for both ODDI-C and MSR. MSR reaches maximum consensus in around 25 seconds while ODDI-C reaches consensus in almost 30 seconds. This trend reverses for higher numbers of disruptors. In these cases, ODDI-C's caution is rewarded as the disruption are observed to have less impact than on MSR's convergence.

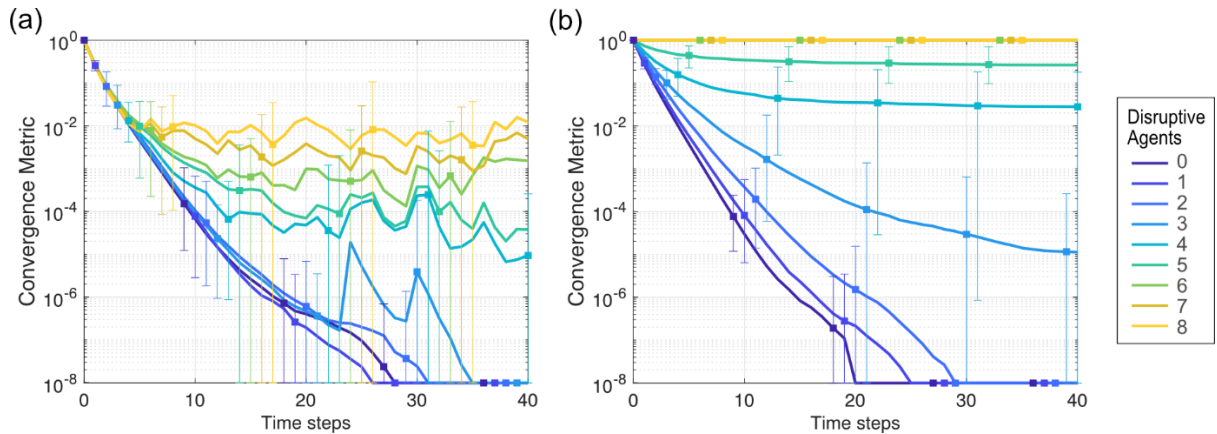

*Figure 5: Comparing the convergence metrics of the ODDI-C (a) and MSR (b) Algorithms.*

### Discussion

Numerical analysis of performance confirms that the ODDI-C Algorithm satisfies its design requirements, meaning the Termination, Agreement, and Weak Validity properties, whilst outperforming MSR when looking at simulation convergence metrics. Experiment 1, through its simulation of the 7-node and 15-node networks (Figure 2 (a) and Figure 3 (a) respectively), validates the ODDI-C algorithm's Weak Validity property. This is further corroborated by Experiment 3's 0-Disruptor test which sees rapid consensus among all nodes. Experiment 1 also demonstrates ODDI-C's adaptability to various types of disruptive behaviour. As shown across both scenarios including disruptive nodes, the algorithm minimises the influence of disruptors. Compared to the mean-based algorithm case, ODDI-C is observed to effectively dampen disruptive instabilities while ensuring convergence of compliant nodes.

Experiments 2 and 3 validate ODDI-C's Termination and Agreement Properties. In addition, they display ODDI-C's adaptability and its advantages compared to MSR, other than the already significant ability to operate without knowledge of the number of disruptive nodes in the network. In both experiments, MSR's trends show a near-constant convergence of compliant nodes until convergence plateaus are reached (see both Figure 4 (b) and Figure 5 (b)). This behaviour is expected from MSR's fixed filtering method. If the number of disruptors is known, filtering out twice that number of the most extreme values received will result in a steady convergence, unaffected by disruptive patterns. This is not the case for ODDI-C. The algorithm filters a non-constant number of values, each time adapting to the values received. As a result, filtering and convergence vary for each node across time steps and runs, as noted by ODDI-C's less constant rate of convergence in both experiments.

In experiment 2, the adaptability is illustrated when assessing both MSR's and ODDI-C's behaviours at high in-degrees. For MSR, increasing connectivity significantly improves the gradient of convergence, in a steady and consistent manner (see Figure 4 (b)). MSR's performance nonetheless remains limited by the amount of filtering being performed. Experiment 2 deals with 5 fixed disruptors, and thus up to 10 values will be filtered out (the 5 highest and lowest values received). In most cases, fewer values will be removed as only values strictly more extreme than the node's own values are removed. Despite this precaution, a vast number of non-malicious nodes will be filtered out, slowing consensus. This is not the case for ODDI-C whose filtering is dynamic and for which increasing connectivity drastically improves performance (see Figure 4 (a)). This improvement in convergence is the result of nodes being provided with more data points/samples from which to create their data distributions and from which to calculate z-scores for filtering (see Section: Algorithm Rules for details of the algorithm's filtering). Increasing connectivity refines the filter, resulting in a faster convergence. While ODDI-C displays high convergence trends, especially when boosted by high in-degrees, it is worth acknowledging a by-product of this adaptability: when exhibiting non-extreme values, disruptive nodes may not be caught by the dynamic filtering and may thus impact the final convergence value. This by-product is observed when looking at ODDI-C's trends for relatively high in-degrees ($V^{in} = 7$) for which oscillations or "wobbles" are observed starting from time step 25. This behaviour, which stems from disrupting sine wave oscillations, emerges only in the ODDI-C simulations. Note, MSR's faces the same disruptors but is also less converged. This variation is the results of distributed consensus' dynamism, specifically, the fact that compliant nodes are always accounting for other nodes' values when creating their filters. Thus, in cases where disruptor nodes' values come close to connected nodes' medians, their disrupting values will influence compliant node's update, making them waver. This characteristic, the wobbling of compliant nodes' convergence, nonetheless remains small (see the log-scale used for the convergence metric).

ODDI-C's adaptability is similarly observed in Experiment 3's performance difference between MSR and ODDI-C. Specifically, for low numbers of disruptors. When dealing with no or low numbers of disruptions, the MSR acts like a classical mean-based consensus, removing none (or very few) of its neighbours. ODDI-C meanwhile, still filters in-coming values. As such, nodes will be accepting a reduced number of state values compared to the MSR, slowing down convergence (see Figure 5). Despite this slight drawback, ODDI-C will still reach convergence; this vigilance is proven to be an advantage whenever disruptions are involved in any scenario. Against low numbers of disruptions, the algorithm's dynamism avoids over-filtering. Unlike the MSR, ODDI-C is therefore able to reach very high levels of convergence without plateauing when presented with disruptions (similar to the patterns observed in Figure 4).

Finally, ODDI-C's adaptability is what allows it to provide a level of convergence when dealing with high ratios of disruptors to connectivity (batches with low connectivity in Figure 4 and high number of disruptors in Figure 5). The higher this ratio, the poorer the convergence, although this remains better than MSR's complete lack of consensus. On MSR's part, this is the result of filtering out all values. For ODDI-C, this stems from the calculation of a median-based z-score. That is, the fewer in-neighbours a node has, the less data points it will have to form a distribution of received values. Z-scores calculated will be coarser and less accurate, with the node's own z-score (which acts as filter discriminant) being more easily influenced by extreme nodes. High ratios of disruptor to connectivity are also much more likely to see one node being overwhelmed by disruptive nodes, and in turn acting disruptive to its own out-neighbours. Nevertheless, it is interesting to note that high levels of performance can still be reached for specific networks (see error bars) even in cases of low connectivities.

In summary, ODDI-C outperforms MSR thanks to its adaptability, despite lacking knowledge of the topology or number of disruptive agents. With its dynamic filtering approach, ODDI-C effectively filters disruptions for a range of disruptor number to connectivity ratios, demonstrating high levels of disruption-resilience. ODDI-C makes the most of increased connectivity and information when its nodes receive it, more so than MSR, as well as continuing to function in runs facing high disruptor ratios which overwhelm MSR. ODDI-C's social dynamics derived resilience is noted across all three experiments, confirming the algorithm as disruption-tolerant and validating its consensus properties across a range of scenarios.

## Conclusion

The mechanics of social opinion dynamics can be used to engineer efficient disruption-tolerant consensus algorithms in fixed networks requiring no knowledge of either network topology or the number of disruptive agents. By adapting the mechanics of tolerance-based opinion diffusion, the Opinion Dynamics-inspired Disruption-tolerant Consensus (ODDI-C) algorithm effectively dampens the instabilities generated by disruptor agents. By curbing the impact of disruptors, shown to affect mean-based algorithm scenarios, the ODDI-C Algorithm enables high levels of convergence to be reached and confirms itself to be disruption-tolerance. The algorithm presented outperforms the classical MSR Method, reaching a tighter set of converged values for the same connectivity in addition to being more robust to the influence of disruptor nodes. The ODDI-C Algorithm enables for better consensus in spite of the presence of disruptive agents and less knowledge than that required by MSR.

## Declarations

### Availability of data and materials

The source code supporting the conclusions of this article is available in the ODDI-C Algorithm GitHub repository, https://github.com/Agathe18474/ODDI-C-Algorithm/tree/v1.0.

Project Name: ODDI-C Algorithm.

Archived version at time of the publication: https://zenodo.org/doi/10.5281/zenodo.10650711

Operating system: Platform independent.

Programming Language: MATLAB 2023b.

Any restrictions to use by non-academics: MATLAB licence needed.

### Competing interests

Not applicable.

### Funding

This project is funded by the Air Force Office of Scientific Research (AFOSR), Grant number FA8655-22-1-7033.

**Authors' contributions**

Conceptualization and methodology, A.B., R.C., C.L., and M.M.; software and numerical experiments, A.B.; Interpretation, A.B., R.C., and M.M.; writing, review, and editing, A.B., R.C., and M.M.; funding acquisition, M.M, and R.C. All authors have read and agreed to the published version of the manuscript.

**Acknowledgements**

Not applicable.